\documentclass[sigconf, 9pt, nonacm]{acmart}

\usepackage{listings}
\usepackage{commath}
\usepackage{graphicx}

\usepackage{amsmath}
\usepackage[utf8]{inputenc}
\usepackage{xcolor}
\usepackage[scriptsize,tight]{subfigure}
\usepackage{diagbox}
\usepackage{algorithm}
\usepackage{algorithmic}
\usepackage{tabularx}
\usepackage{amsmath}
\usepackage{amsfonts}
\usepackage{enumitem}
\usepackage{bm}
\usepackage{multirow}
\usepackage{geometry}
\long\def\comment#1{}

\newcommand{\name}{MCT}

\newcommand{\df}{\mathrm{d}}

\newcommand{\sssec}[1]{\vspace*{0.02in}\noindent\textbf{#1}}

\newcommand{\change}[1]{{\color{black} #1}}

\def\l@subsubsubsection{\@dottedtocline{4}{4.8em}{4.2em}}

\title{Towards Seeing Bones at Radio Frequency}

\author{Yiwen Song}
\email{yiwens2@andrew.cmu.edu}
\affiliation{
  \institution{Carnegie Mellon University}
  \country{USA}}

\author{Hongyang Li}
\affiliation{
  \institution{Carnegie Mellon University \& UW-Madison}
  \country{USA}}

\author{Kuang Yuan}
\affiliation{
  \institution{Carnegie Mellon University}
  \country{USA}}

\author{Ran Bi}
\affiliation{
  \institution{Carnegie Mellon University \& Rice University}
  \country{USA}}

\author{Swarun Kumar}
\email{swarun@cmu.edu}
\affiliation{
  \institution{Carnegie Mellon University}
  \country{USA}}

\begin{document}


\begin{abstract}
Wireless sensing literature has long aspired to achieve X-ray-like vision at radio frequencies. Yet, state-of-the-art wireless sensing literature has yet to generate the archetypal X-ray image: one of the bones beneath flesh. In this paper, we explore \name, a penetration-based RF-imaging system for imaging bones at mm-resolution, one that significantly exceeds prior penetration-based RF imaging literature. Indeed the long wavelength, significant attenuation and complex diffraction that occur as RF propagates through flesh, have long limited imaging resolution (to several centimeters at best). We address these concerns through a novel penetration-based synthetic aperture algorithm, coupled with a learning-based pipeline to correct for diffraction-induced artifacts. A detailed evaluation of meat models demonstrates a resolution improvement from sub-decimeter to sub-centimeter over prior art in RF penetrative imaging.


\end{abstract}

\maketitle


\section{Introduction}

\change{
Wireless sensing and imaging at Radio Frequency (RF) is well-known for its ability to sense through occlusions.
Past works have demonstrated through-fog and through-wall sensing and imaging using millimeter~\cite{Prabhakara_2023,guan2020through,deng2022geryon} and Wi-Fi signals~\cite{mostofi2012cooperative,adib2013see,zhang2023understanding}.
This paper embraces the the ability of radio frequency signals to penetrate through objects by asking a unique question -- ``How can we image the cross section of bones inside flesh using radio frequency?'' 
Radio frequency wireless sensing has seen massive advances for vital sign sensing, body posture imaging and even body silhouette monitoring, which achieves high resolution by measuring and processing a reflective signal from body surface or shallow tissues~\cite{liu2015tracking,ali2021goodness,jiang2020towards,ji2023construct,liu2024real,li2018wifit}. 
Much of this work either uses communication radios, custom or commodity radars operating at RF, at frequencies that are non-ionizing and therefore dramatically safer than X-Rays in sensing through body. 
Yet, none of the prior work on RF penetrative imaging can image bones, and thereby help study bone health without the radiation risks of X-Rays. 
Instead, these works largely rely on measuring the reflections off the surface of objects.
For example, prior art on vital sign monitoring such as breathing or heart rate looks for surface vibrations induced by these phenomena rather than imaging the lung or heart. 
Even with the presence of occlusions, past works rely on filtering out the reflections from walls and process the reflective signal from the object.
}

\change{The reason for RF reflective imaging's inability to image bones is that the tissue, mainly constituted of water~\cite{adolph1933metabolism}, is highly absorptive ($\sim$20~dB for 5~GHz).
Therefore, the reflective signal from bones, after the penetration inside flesh, will be 40~dB less than the direct reflection from the skin surface, which will be hard to detect.
With the penetrative imaging, it is possible to measure the difference in penetration loss between different tissue materials to distinguish them deep inside body.
However, the heavily weakened RF signals that do penetrate through the body interact with the body in complex ways (owing to diffraction, material-dependent attenuation, etc.), rendering imaging challenging. } 
The few papers that have performed penetrative imaging using RF \change{have largely focused on large objects (ships~\cite{noviello2024multiple,lu2023improved}, trees~\cite{boero2018microwave}, building structures~\cite{mostofi2012cooperative} } and landscapes~\cite{chen2018landscape,sundaram2022microwave}) or on body organs with extremely poor (decimeter-level) resolution~\cite{alvarez2018synthetic, semenov2007microwave, semenov1996microwave, alkhodari2022using, alqadami2018wearable}. 
To the best of our knowledge, no prior system has been demonstrated sub cm-scale imaging of bone tissue using RF. 

This paper presents \name, a microwave computed tomography system operating at 5-6~GHz RF that images 2D cross sections of bones through flesh. 
\change{
Compared to X-ray and MRI, \name\ offers the advantage of avoiding ionizing radiation while maintaining a much lower operating cost, which allows much more frequent imaging, at the cost of less resolution.
\name\ acts as a feasibility study to show a fully non-contact deep-body high-resolution imaging system with radio frequency.
}
We conduct our evaluation on meat models rather than human tissue owing to the ease of ground truth.
\change{By slicing through the meat, we are able to measure the propagation constants without X-ray imaging and capture accurate ground truth values. }
\name\ functions by directing focused RF beams around an object of interest from multiple perspectives and detecting weak RF signal receptions to reconstruct a 2D cross-section of tissue. 
Our rigorous evaluation of \name\ on meat models demonstrates promising imaging accuracy and resolution.
The evaluation further shows \name's potential in studying key bone health parameters such as density, location, and bone-muscle unit sensing.

\begin{figure}
    \centering
    \includegraphics[width=0.9\linewidth]{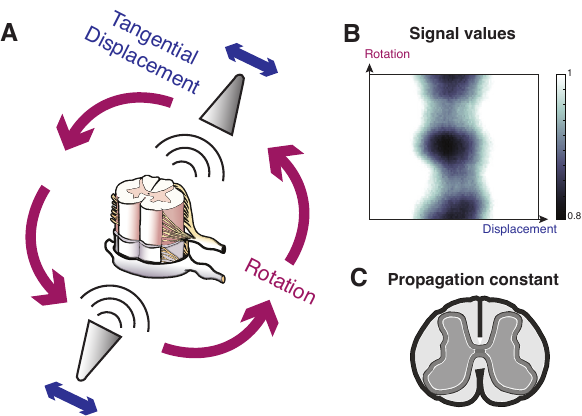}
    \vspace*{-0.05in}
    \caption{Introduction: (A) \name\ includes two antennas with rotation and tangential displacement. (B) The received signal with regard to antenna rotation and displacement. (C) The propagation constant inferred by the collected signal values.}
    \label{fig:intro}
    \vspace*{-0.1in}
\end{figure}

At a high level, \name\ computes and images the RF propagation constants inside an object. 
\change{Assume the RF propagates through a meat model and a bone model of the same size, the difference in material property creates a difference in the RF propagation loss and phase change.}
\change{To find where the material is located, the working principle of microwave CT is to create diversity in the propagation path through the object, therefore obtaining the signal propagating through different regions. }
In practice, we create a single transmitter-receiver pair positioned on opposite sides of the object, which captures signal changes from diverse perspectives as they systematically move around the object.
We note that the system requires the object to be static during the scanning process, much like the X-ray-based CT system.
We characterize the systematic movements of the transmitter-receiver pair by a combination of rotation or tangential displacement, shown in Figure~\ref{fig:intro}A. 
A captured signal profile shown in Figure~\ref{fig:intro}B visualizes the signal attenuation at each rotation and displacement position of the antenna pair. 
Based on the captured signal profile, \name's algorithm further estimates the signal absorption inside the object shown in Figure~\ref{fig:intro}C.

There are two main challenges in constructing a high-resolution propagation constant map of the object, due to the unique properties of RF compared to X-rays.
First, unlike optical diodes, RF transceivers have a large aperture. 
The large aperture causes poor resolution as the transceivers measure signal attenuation across a large area.
\name\ proposes a new sub-wavelength penetration-based synthetic aperture algorithm to reduce the effective aperture of transceivers.
Although the concept of synthetic aperture has been widely used in RF-reflective systems such as radar~\cite{curlander1991synthetic,franceschetti2018synthetic,lai2024enabling, xu2022mask} and backscatter~\cite{adib2013see,wu2023enabling,colone2013wifi}, synthetic aperture systems for RF penetration have not been widely studied.
Past synthetic aperture systems largely target wavelength-scale imaging~\cite{gao2021mimo,li2022low}.
\name\ proposes a synthetic aperture algorithm that achieves sub-wavelength scale imaging, pushing the limit of the concept of synthetic apertures.
The basic idea of \name's synthetic aperture design is as follows.
Since the RF transceivers have a large aperture, i.e., a wide beam crossing the object, it has a poor resolution since we can not determine which part of the object accounts for the signal change.
By creating a small displacement of the transceiver pairs, although the physical aperture is not changed, the signal changes due to the difference between the old beam and the new beam.
Therefore, the difference constitutes a virtual narrow beam with a tiny synthetic aperture, allowing \name\ to image with a resolution that is far less than the operating wavelength.
Solving the propagation constant from the synthetic apertures is not trivial, since the RF waves experience both amplitude and phase change during propagation.
We propose a novel formulation and phase-retrieval algorithms to solve a propagation constant map inside the object.

Second, even with the penetration-based synthetic aperture method, the solved propagation constant map is not accurate. 
This is because RF penetration can not be simply characterized by ray optics.
Due to the large wavelength of the RF signal, diffraction at medium interfaces --- particularly between object and air --- creates spurious artifacts in our recovered images, including Poisson spots at the center and diffraction fringes beyond edges. 
While diffraction can be estimated through finite-element methods (FEM) for known objects, removing these artifacts (the inverse problem) is computationally intractable using traditional optimization approaches. 
Instead, we take a data-driven approach by using a U-Net neural network structure that learns to map from diffraction-contaminated images to clean reconstructions. The encoder-decoder architecture with skip connections in U-Net is effective for this task as it captures multi-scale features while preserving spatial information critical for artifact removal. We further enhance reconstruction by exploiting frequency diversity, based on the observation that diffraction patterns vary across frequencies while the underlying object structure remains constant. Specifically, we fuse the information across different frequencies by averaging the embeddings encoded from different subbands in 5-6 GHz, at the bottleneck and skip connections of the U-Net. Finally, we design a data augmentation technique by using various combinations of subbands as input during training to improve the robustness of our model.

We implement \name\ factoring in several practical considerations, including frequency selection, antenna directivity and aperture, system size, and software/hardware selection.
\name\ is evaluated using chops of pork forearms and ribs with different bone distributions. 
We use a clustering algorithm to classify bones and other fleshes.
The evaluation result shows a 0.86 SSIM and 68~dB PSNR can be acquired in accessing the similarities between ground truth and recovered propagation constants.
We further evaluate the clustered result, showing a 1~mm average resolution in locating bones, and a 93.4\% segmentation accuracy of bones and flesh.

\sssec{Contributions: } Our contributions are as follows.
    \vspace*{-0.05in}
\begin{itemize}[leftmargin=*]
    \item \name\ is the first microwave system that achieves sub-centimeter-level in-flesh imaging. \name\ achieves this by proposing a novel sub-wavelength penetration-based synthetic-aperture algorithm for recovering propagation constants.
    \item \name\ proposes a model to characterize the artifacts brought about by microwave diffraction around the object, and develop a neural network with multi-frequency fusion to remove the diffraction artifacts.
    \item \name\ validates the proposed algorithm with a variety of objects. Evaluation results show a significant 0.3 SSIM improvement and 48~dB PSNR improvement compared to prior work, improving the resolution from sub-decimeter to sub-centimeter.
\end{itemize}

\vspace*{-0.05in}\noindent \textbf{Limitations: } We wish to highlight 3 important limitations of \name: (1) \name\ delivers only 2D cross-sectional images, rather than full-blown 3D images; 
(2) Same as a regular CT scan, \name's implementation requires collecting a large amount of I/Q data with mechanical scanning and is more time-consuming than a single-shot image; 
(3) \name\ has been evaluated on meat models owing to the ease of ground truth (collecting cross-sectional images) -- we therefore do not perform human trials. 
While we leave addressing these limitations for future work, we note that \name\ nevertheless presents an important step forward in through-flesh bone imaging using RF. 

\begin{figure*}[hbt!]
    \centering
    \includegraphics[width=0.9\linewidth]{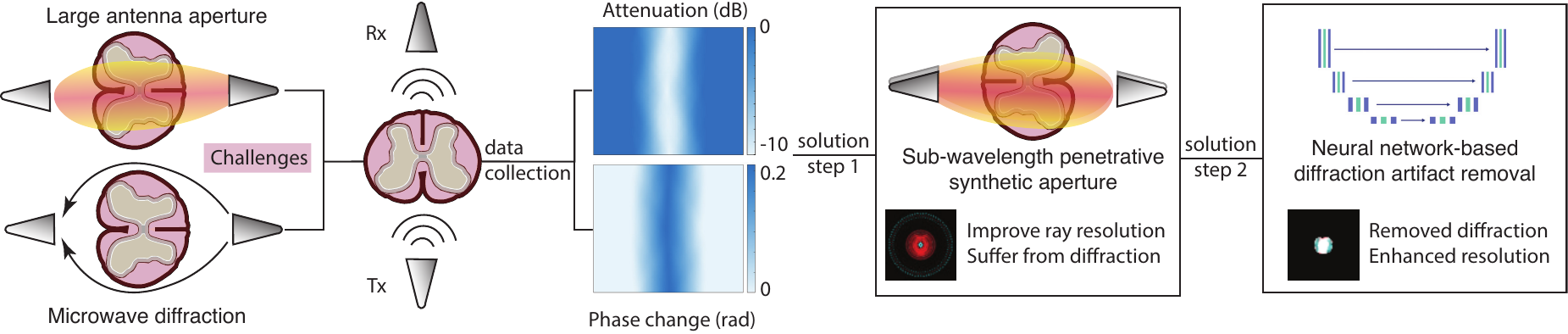}\vspace{-5pt}
    \caption{System overview. The key contributions of this paper include: (1) To deal with large antenna aperture at microwave frequency, we propose a sub-wavelength penetrative synthetic aperture algorithm to create synthetic needle-like apertures, thus improving resolution. (2) To deal with microwave diffraction around the object, we create a data-driven neural network to suppress the diffraction artifact.}\vspace{-5pt}
    \label{fig:overview}
    \Description{System overview}
\end{figure*}

\begin{table*}[]
\begin{tabular}{c|c|c|c|c|c}
\hline
\textbf{Method}             & \textbf{\begin{tabular}[c]{@{}c@{}}Indentify\\ materials\end{tabular}} & \textbf{\begin{tabular}[c]{@{}c@{}}Image\\ external shape\end{tabular}} & \textbf{\begin{tabular}[c]{@{}c@{}}Image\\ internal shape\end{tabular}} & \textbf{Resolution} & \textbf{Non-Ionizing}\\
\hline\hline
X-Ray                   & Yes                          & Yes                           & Yes                           & Very high (sub-mm) & No  \\
Wireless Material Sensing~\cite{wu2020msense,zhang2019feasibility,ren2020liquid,yan2024wi}   & Yes                          & No                            & No                            & N/A & Yes                 \\
WiFi Imaging~\cite{wang2019person,li2020wi,karanam20173d}                & No                           & Yes                           & No                            & Medium (cm-dm) & Yes      \\
mmWave Imaging~\cite{aladsani2019leveraging,Prabhakara_2023}              & No                           & Yes                           & No                            & High (sub-cm) & Yes       \\
Microwave Remote Tomography~\cite{noviello2024multiple,lu2023improved} & Yes                          & Yes                           & No                            & Low (m) & Yes             \\
Microwave In-body Imaging~\cite{mostofi2012cooperative}   & Yes                          & Yes/No                        & Yes                           & Medium (cm-dm) & Yes     \\
MCT (Ours)                  & Yes                          & Yes                           & Yes                           & High (sub-cm) & Yes    
\\
\hline
\end{tabular}
\caption{A comparison of state-of-the-art wireless sensing and imaging works. }
\vspace{-15pt}
\label{tbl:related_work}
\end{table*}

\section{Related Work}

Table~\ref{tbl:related_work} summarizes a comparison between our work with state-of-the-art works.

\noindent\textbf{Imaging at Radio Frequency: } 
\change{
Imaging objects at radio frequency utilizes bands including WiFi~\cite{wang2019person,li2020wi}, millimeter wave~\cite{aladsani2019leveraging,Prabhakara_2023},  UWB~\cite{zhuge2010sparse,yarovoy2007uwb}, etc. 
The works mostly rely on reflective signals from objects, and thus only image the surface mesh of the objects.
Recently through-wall imaging also utilizes reflective signals, but filters out the first reflection from walls to enable high-quality images~\cite{karanam20173d,huang2009uwb}.
}
However, they are still not capable of imaging deep inside body.
Alternatively, microwave tomography techniques utilizes penetrative wireless signals.
Most works focus on using multiple static transceiver pairs to image large objects such as ships~\cite{noviello2024multiple,lu2023improved}, trees~\cite{boero2018microwave}, and landscapes~\cite{chen2018landscape,sundaram2022microwave}, etc., compared to which the microwave wavelength is negligible.
Simulations have shown the feasibility of microwave imaging inside human tissues~\cite{fedeli2021microwave,hosseinzadegan2021discrete,meaney2022contemporary,cannata2024microwave,lalitha2022non}.
However, moving from simulation to real-world systems leaves practical challenges.
Previous experimental studies have included studying human head models~\cite{lu2022three,guo2024calibrated, islam2022portable}, items buried under soils~\cite{catapano2023contactless,esposito2025quantitative,catapano2022gpr,catapano2023ground,alkhodari2021monitoring}, bone models~\cite{amin2022experimental,santos2022feasibility,cannata2024exploring}, breast~\cite{1528780}, and security screening~\cite{advanced_micro}.
However, these prior results failed to provide high-quality images of the object deep inside body. 
The main challenge is that the microwave channel is far more complex than optical channels~\cite{6305002,alsawaftah2022microwave}.
Limited resolutions can be achieved from past works attempting microwave imaging of human tissue models such as bones, breasts and brains~\cite{semenov2002three,semenov2005microwave,semenov2016electromagnetic,Ruvio_1, Elahi_1,mostofi2012cooperative}
They do not reflect the internal structures of the object, usually blurring the edges of the internal structures as circles or have many artifacts. 
The attempts to image the edges of the internal structures more clearly overlook the uneven thickness of flesh~\cite{8620220}, or require unrealistic testing environments, such as immersing the object in coupling liquid to minimize diffraction effects~\cite{clinical_prototype, 5872561,1528780}, or relying on some prior knowledge ~\cite{6242390}. As past solutions heavily rely on small revisions of the X-ray tomography, a lack of consideration for complex signals, large microwave apertures, and diffraction, resulting in poor resolution, especially incapability of detecting clear object edges.
Past work demonstrates the potential for improving image quality by improving the signal processing pipeline.

\noindent\textbf{Material Sensing at Radio Frequency: }
Radio-frequency senses dielectric materials by sensing the signal change caused by different dielectric properties of the material, mainly the dielectric constant.
Most systems focus on sensing an amplitude and phase change in the reflected signals~\cite{wu2020msense,zhang2019feasibility,ren2020liquid,yan2024wi}.
They usually require a homogeneous uniform material to be sensed.
Signals can also come from reflection from the deeper object, but is much fainter than the surface reflection.
Recent systems have successfully leveraged the low SNR signal changes for more complicated sensing tasks~\cite{sun2024rf,afzal2023agritera,karmakar2023meta}.
However, they need a reference object or a long-term measurement for system calibration.
\name\ uses more measurements to get more information inside the object, and thus can image without a calibration step and prior knowledge of the object.

\noindent\textbf{Health sensing: } Past work on wireless health sensing focuses on sensing through a reflected signal from body, leveraging wireless technologies including WiFi~\cite{wang2022wi,jiang2020towards}, acoustics~\cite{SpiroSonic, EarHealth, DF-Sense}, microwave~\cite{Emotion_recognition, Cardiac_Scan}, RFID~\cite{ZEROECG} and millimeter-wave radar~\cite{Contactless_seismocardiography}.
Prior systems for sensing breathing, cardiovascular activity, or ear health classify signals based on skin vibrations from heartbeat, blood flow, or breath, as well as echoes from ear canal or eardrum vibrations, rather than directly imaging or sensing the organs.
Other solutions have focused on using radio-frequency penetration inside the body for sensing~\cite{tao2023magnetic,ma2024rf} and power delivery~\cite{fan2020towards}.
In-body impedance tomography~\cite{stefaniak2024classification, baran2022bets, wojcik2021diagnosing} uses a similar principle as microwave tomography but needs an array of contact electrodes to be attached to the skin.


\section{Overview}

Figure~\ref{fig:overview} shows our system overview. We first highlight the two major challenges in \name\ --- large antenna aperture and microwave diffraction --- by characterizing the differences between microwave CT and X-ray CT through comprehensive modeling~(Sec.~\ref{sec:modeling}). 
We then propose a two-step approach in Sec.~\ref{sec:solution} to mitigate the two challenges accordingly. 
In the first step (Sec.~\ref{subsec:sar}), we develop a synthetic aperture optimization algorithm through antenna movement, which creates a sub-ray model to utilize relative displacement with the antenna and the object.
However, the output of the first step is still disturbed by diffraction artifacts.
Sec.~\ref{subsec:unet} developed a neural network to remove diffraction features.
We develop a U-Net-based neural network with multi-frequency fusion to capture real-world diffraction features and enhance image recovery performance.
Finally, we present the implementation details in practical deployment settings (Sec.~\ref{sec:implementation}) and the evaluation result~(Sec.~\ref{sec:result}).

\section{Understanding CT at RF vs. X-Ray}\label{sec:modeling}

\subsection{A Primer on Traditional X-Ray CT}

\begin{figure*}
    \centering
    \includegraphics[width=0.85\linewidth]{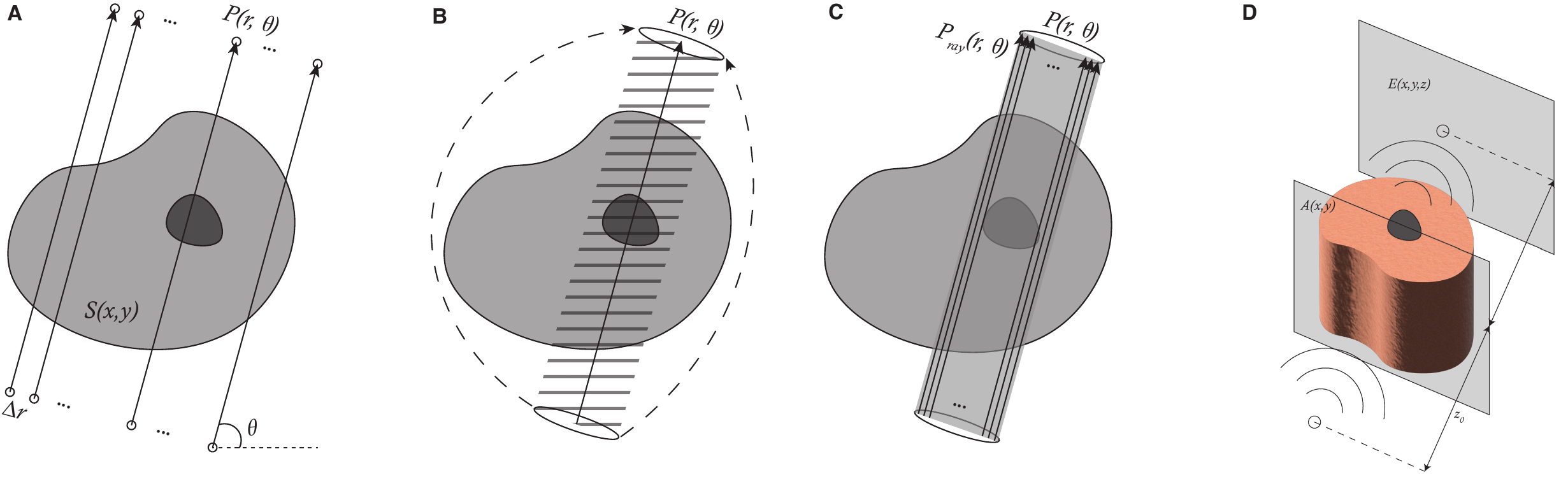}\vspace{-10pt}
    \caption{Modeling of (A) traditional CT and (B) microwave CT. The main difference in modeling includes large ray width that can not be neglected (dashed area) and diffraction of the objects (dashed line). (C) Synthetic aperture formulation, where a wide ray is divided into a sum of several thin sub-rays. (D) Diffraction formulation, where the cross-section of the object is considered to calculate the electric field projection on the receiver plane.}\vspace{-10pt}
    \label{fig:ct-modeling}
    \Description{ct modeling}
\end{figure*}

We first present a brief primer on how traditional X-ray CT works and why the method of X-ray CT can not be directly applied to the microwave scenario.
Traditional X-ray CT measures the signal attenuation across different spatial regions to compute an absorption spectrum inside the object. 
Specifically, consider an object to be imaged, where the absorption rate of a point $(x,y)$ is denoted by $S(x,y)\in\mathbb{R}^+$, as shown in Figure~\ref{fig:ct-modeling}A. 
A parallel pair of transmitters and receivers is placed on both sides of the object. 
The elements are separated by $\Delta r$. 
There are $T$ rotation rounds, where in each round the array is rotated to a new angle $\theta$ relative to the object. 
A transceiver pair in one round transmits a signal level of 1 (normalized) and receives a signal level of $P(r, \theta)\in (0,1)$, where $r$ represents its spatial offset and $\theta$ represents its rotary angle. 
X-ray has a short wavelength and thus ray optics can be adapted, where the attenuation of the signal level observed in $P(r,\theta)$ is constituted of the exponential attenuation through each spatial element, i.e.:
\begin{align}\label{eq:traditional_ct_formulation}
    \begin{split}
    P(r, \theta)&=\exp\left(-\int_{L}S(x,y)\df l\right),\\
    L&: -x\sin\theta + y\cos\theta=r.
    \end{split}
\end{align}

CT solves $S(x,y)$ from the observed $P(r,\theta)$ in the discrete domain. Specifically, the following optimization problem is solved.
\begin{align}\label{eq:traditioal_ct_optimization}
    \begin{split}
        \min_{S(x,y)}\quad & \sum_{r, \theta}\quad \left(\log P(r, \theta)+\sum_{d(x,y|L)\leq\Delta r/2}S(x,y)\right)^2, \\
        \mathrm{s.t.}\quad & S(x,y)\geq 0,
    \end{split}
\end{align}

where $d(x,y|L)$ denotes the distance from point $(x,y)$ to ray $L$. 
To solve it, denote $Q(r,\theta)=-\log P(r,\theta)$. 
Let $\bm{Q}$ be a row-vectorized version of $Q(r,\theta)$, and $\bm{S}$ be a row-vectorized version of $S(x,y)$. 
The sum term in Eq.~\eqref{eq:traditioal_ct_optimization} of $S(x,y)$ can then be reformulated as $\bm{R}(r, \theta)^{\intercal}\bm{S}$, where each element of $\bm{R}(r, \theta)$ denotes the result of the indicator function $\mathbb{I}[d(x,y|l)\leq\Delta r/2]$. The vector function $\bm{R}(r, \theta)$ can be further concatenated into a matrix $\bm{R}$ using the same vectorizing index of $\bm{Q}$. The absorption spectrum $\bm{S}$ inside the object can thus be solved by the solving following linearly-constrained quadratic programming (LCQP) problem.
\begin{align}\label{eq:traditioal_ct_lcqp}
    \begin{split}
        \min_{\bm{S}}\quad & \left\lVert\bm{Q}-\bm{R}^{\intercal}\bm{S}\right\rVert_2^2\\
        \mathrm{s.t.}\quad & \bm{S}\geq 0, \mathrm{elementwise}.
    \end{split}
\end{align}

\subsection{Building a Microwave CT Model}

\textbf{Difference with X-ray CT:} 
The characteristic of electromagnetic wave propagation through human-sized objects in microwave frequency is significantly different from X-ray. 
We demonstrate a modeling in Figure~\ref{fig:ct-modeling}B. 
First, considering the non-negligible aperture of microwave transceivers, the Tx-Rx pair can not be formulated as point transceivers. Instead, it has a ray width of $z$. 
Moreover, due to the ray width, the phase change in wave propagation can not be neglected.

\noindent\textbf{Plane wave propagation:} 
The concept of plane wave propagation is the physical foundation of our system.
Consider a 1D case, for a plane wave starting at position 0, propagating through a homogenous medium with a complex propagation constant $\alpha+j\beta$, where $j=\sqrt{-1}$, the phasor of the electric field $\hat{E}(z)$ at position $z$ can be calculated as follows.

\begin{equation}
    \hat{E}(z)=\hat{E}(0)\exp(-\alpha z-j\beta z).
\end{equation}

The equation states two important properties of a medium.
First, the electric field intensity (signal strength) decreases exponentially when transmitting through a medium. 
The decay factor $\alpha$ constitutes the real part of the propagation constant, and is larger than 0.
Second, the phase decreases along the propagation direction due to the time delay from position $z$ to position 0.
The phase factor $\beta$ constitutes the imaginary part of the propagation constant, which is inversely proportional to the speed of light in the medium.

\noindent\textbf{Wave propagation through object:} 
Due to Huygens–Fresnel principle, each point on the wavefront between the transceiver can be viewed as an independent wave source, where the sum of the waves from the sources constitute the received signal.
As shown in Figure~\ref{fig:ct-modeling}C, we divide the transmitter and the receiver into a lot of tiny transmitter-receiver pairs, and calculate the signal through each transceiver pair. 
The waves can add together constructively or destructively, depending on their phase.
Each point of the object has a complex attenuation constant $S(x,y)\in\mathbb{C}$, corresponding to the complex propagation constant. 
The ray signal $P_{\mathrm{ray}}(r, \theta)\in\mathbb{C}$ from Tx to Rx can be derived as follows.

\begin{align}\label{eq:microwave_ct_ray_model}
    \begin{split}
        P_{\mathrm{ray}}(r, \theta) & = \int_{Z}\exp\left(-\int_{L_z} S(x,y)\df l_z\right)\df z, \\
        Z & : x\cos\theta - y\sin\theta+r\sin2\theta=0,\\
           & x\in\left(-(r-\frac{z}{2})\sin\theta, -(r+\frac{z}{2})\sin\theta\right),\\
        L_z & : -x\sin\theta + y\cos\theta=r_z,\\
        r_z & = r+\Delta z, -\frac{z}{2}\leq\Delta z\leq\frac{z}{2}.
    \end{split}
\end{align}

In the equation, the received signal is a sum of the small sub-rays that have a very small aperture, which is defined by the step of displacement of the antenna (0.5~cm in our case).
Essentially, the difference between the two adjacent rays can represented as a sub-ray.

\noindent\textbf{Wave diffraction around object:} We then consider the effect of diffraction. 
To determine a model for diffraction, we must first compute the Fresnel number $F$ of our system. 
It is given by $F=\frac{A}{L\lambda}$, where $A$ is the size of the object cross-section, $L$ is the distance from the object to the receiver antenna, and $\lambda$ is the wavelength of the 5-6~GHz signal, which is between 5~cm to 6~cm. 
We use a system (See Sec.~\ref{sec:implementation}) with Fresnel number around 1. 
Therefore, we characterize the system using Fresnel diffraction. 
An electric field induced by Fresnel diffraction can be calculated by solving the Helmholtz equation in the space of the object and transceivers. 
However, acquiring a numerical solution for the Helmholtz equation of a complex object requires heavy computation through finite element methods.
Therefore, we simplify the model of the object to acquire an approximated diffraction pattern.
To simplify the object, we omit the diffraction caused inside the object, since the diffraction inside the object is much lighter than the wave penetration through the object. 
Therefore, we model the object as a perfect absorber.
Furthermore, the diffraction is mainly caused by the maximum vertical cross-section of the object.
Considering the two simplifications, the diffraction of the object can be modeled as a 2D curtain between the transmitter and the receiver. 
Assume a curtain defined by $\mathcal{A}_{x,y}$ on the $xOy$ plane.
The transceivers are of distance $z_0$ to the curtain, shown in Figure~\ref{fig:ct-modeling} whose positions are denoted by $(0,0,-z_0)$ and $(0,0,z_0)$.
We have:
\begin{equation}
    E(x,y,0)=0, \forall(x,y)\in \mathcal{A}_{x,y}.
\end{equation}

The electric field intensity at the cross-section plane can be calculated according to the inverse square law. 
With a unit E-field $E_0$, the electric field intensity is calculated through a linear combination of terms of order $\frac{1}{d}, \frac{\lambda}{2\pi d^2}, \frac{\lambda^2}{4\pi^2 d^3}$. 
In our system, we omit the term $\frac{\lambda}{2\pi d}\approx\frac{1}{25}$ since it is much smaller than 1. 
Therefore, we can approximate the electric field intensity at the cross-sectional plane $xOy$ as follows.
\begin{align}\label{eq:e_field_on_cross_section}
\begin{split}
        E(x,y,0) & = \begin{cases}
            0, & (x,y)\in\mathcal{A}_{x,y}\\
            \frac{E_0}{d}\exp(\frac{j2\pi d}{\lambda}), & (x,y)\notin\mathcal{A}_{x,y}
        \end{cases},\\
        d & =\frac{1}{\sqrt{x^2+y^2+z_0^2}},
    \end{split}
\end{align}

where $j=\sqrt{-1}$.
From the electric field intensity on the cross-section plane, we can calculate the electric field intensity on the receiver plane $E(x,y,z_0)$ according to the Fresnel diffraction equation as follows.
\begin{equation}\label{eq:e_field_on_receiver_plane}
    E(x,y,z_0)=\int_{-\infty}^{\infty}\int_{-\infty}^{\infty}E(x',y',0)\frac{z_0e^{\frac{j2d\pi}{\lambda}}}{d^2}(\frac{1}{j\lambda}+\frac{1}{2d\pi})\df x'\df y'.
\end{equation} 

\sssec{Putting it all together: } The signal that is received by the receiver can thus be modeled as the integration of the Poynting vector over the receiving antenna aperture $\mathcal{A}_t$. 
The Poynting vector can be calculated as the square of electric field over the intrinsic impedance of air, which is 376~$\Omega$. 
The integration can be approximated as scalar integration due to the small antenna aperture.
\begin{align}\label{eq:power_diffraction}
    \begin{split}
    \lvert P_{\mathrm{diffract}}\rvert & =\iint_{\mathcal{A}_t}\frac{\lvert E(x,y,z_0)\rvert^2}{376~\Omega}\df s, \\
    \angle P_{\mathrm{diffract}} & = \angle E(x,y,z_0).
    \end{split}
\end{align}

The overall signal power and phase at the receiver can then be written as an affine combination of the power result by the ray models and the power result by the diffraction.
\begin{align}\label{eq:power_combined}
    P(r, \theta) & = \alpha P_{\mathrm{ray}}(r,\theta)+ (1-\alpha) P_{\mathrm{diffract}}(r,\theta),
\end{align}

where the $P_{\mathrm{diffract}}(r,\theta)$ is calculated by Eq.~\eqref{eq:power_diffraction} with the object cross-section $\mathcal{A}_{x,y}(r,\theta)$ computed as the projection of the object on the plane normal to the transmitter-receiver trace, shown in Figure~\ref{fig:ct-modeling}D.


\section{MCT's System Design}\label{sec:solution}

\subsection{Imaging at High Resolution via Penetration-Based Synthetic Aperture}
\label{subsec:sar}


We improve imaging accuracy through a form of penetration-based synthetic aperture. Conceptually, while the beams between \name's transmitter and receiver may be wide, as they rotate around the object, the \textit{difference} between these beams is narrow. In the ray model sense, we consider a broad beam to be composed of multiple narrow rays, and therefore when considering differences between broad beams, only a few narrow rays persist.

\sssec{Penetration-Based Synthetic Aperture Model: } Computationally, we can only solve $S(x,y)$ in the discrete domain. 
To achieve this, we divide $S(x,y)$ into a grid of size $N_x\times N_y$, and the value of each pixel in the grid represents the values of $S(x,y)$. 
We reformulate Eq.~\eqref{eq:microwave_ct_ray_model} into a discrete form for optimization as follows.
\begin{align}\label{eq:microwave_ct_ray_discrete_model}
    \begin{split}
        P_{\mathrm{ray}}(r,\theta) =& \sum_{L_z\in \mathcal{Z}}\exp\left(-\sum_{d(x,y|L_z)\leq\delta}S(x,y)\right),\\
        \mathcal{Z}=&\left\{(x,y): \begin{array}{l}
            -x\sin\theta + y\cos\theta=r_z,\\
            r_z= r\pm 2n\delta, n=0, 1, \cdots
        \end{array}\right\}\\
    \end{split}
\end{align}

Recall that in Eq.~\eqref{eq:microwave_ct_ray_model}, a ray consists of a width defined by $Z$ and the line passing through the object defined by $L_z$, corresponding to the two integrals. 
The discretizing step divides the ray width $Z$ into multiple disjoint small thin rays, with their width denoted by $2\delta$.
The center line of the ray is defined in the set $\mathcal{Z}$. 
Therefore, in the discrete formulation of $P_{\mathrm{ray}}(r,\theta)$, the inner bracket $\sum_{d(x,y|L_z)\leq\delta}S(x,y)$ calculates the sum of the complex attenuation constant $S(x,y)$ in the area that is covered by the sub-ray defined by a center line $L_z$ and width $2\delta$. 
The signal result from the summed attenuation is calculated by taking an exponential.
The small thin rays are summed up as the overall signal.

\sssec{Formulating the Imaging Problem: } We then present an optimization approach to solve $S(x,y)$ from the measurements $P_{\mathrm{ray}}(x,y)$, temporarily treating the signal caused by diffraction as noise.
Following the similar process as we formulate Eq.~\eqref{eq:traditioal_ct_optimization} to Eq.~\eqref{eq:traditioal_ct_lcqp}, we vectorize $P(r,\theta)$ into a complex vector $\bm{P}\in\mathbb{C}^{N_rN_\theta\times1}$, and $S(x,y)$ into a complex vector $\bm{S}\in\mathbb{C}^{N_xN_y\times1}$. 
The two summations can be formulated into two matrix multiplications, denoted by
\begin{equation}
    \bm{P} = \bm{R}_x\exp\left(-\bm{R}_r\bm{S}\right),
\end{equation}
where $\bm{R}_x$ and $\bm{R}_r$ are the summation matrices. 
Each row in $\bm{R}_r$ represents a narrow sub-ray, where the elements of $\bm{R}_r$ are calculated by Siddon's algorithm~\cite{siddon1985fast}, representing if the corresponding value in $\bm{S}$ will be included in the sub-ray.
Similarly, each row in $\bm{R}_x$ represents a ray defined by $(r,\theta)$, and each element is a 0 or 1, representing if the corresponding sub-ray will be included in the ray.
Therefore, the optimization problem can be formulated as follows.
\begin{align}
    \min_{\bm{S}} \quad & \left\lVert\bm{P}-\bm{R}_x\exp\left(-\bm{R}_r\bm{S}\right)\right\rVert_2^2+\mu\left\lVert\bm{S}\right\rVert_2^2,\\
    \mathrm{s.t.} \quad & \mathrm{Re}[\bm{S}]\geq 0,\\
    & \mathrm{Im}[\bm{S}]\geq 0.
\end{align}

Note that a necessary condition for the problem to be solvable is that the number of elements in $\bm{S}$ is smaller than the number of elements in $\bm{P}$, i.e., $N_xN_y\leq N_rN_\theta$.

\sssec{Why do traditional algorithms fail?: } 
We need to solve the complex absorption rate $S(x,y)$ from the signal power obtained through scanning along $r$ and rotating along $\theta$. 
An intuitive way of solving $S(x,y)$ from Eq.~\eqref{eq:power_combined} is by an iterative solver from an appropriate point.
Such iterative algorithms have been proposed for other models for microwave and X-ray tomography, but do not fit the context of our modeling.
There are three reasons underlying the claim.
First, unlike the real problems that old algorithms try to solve, our problem is a complex problem, which involves phase retrieval. 
However, general phase retrieval solutions, such as PhaseLift, are complex in both time and space.
Considering the large number of elements in $S(x,y)$ and $P(r,\theta)$, an algorithm such as PhaseLift can take several days in a highest-end CPU to solve~\cite{candes2013phaselift, bandeira2014phase, li2020phase}.
Second, the calculation of Eq.~\eqref{eq:e_field_on_receiver_plane} and Eq.~\eqref{eq:power_diffraction} takes time due to using a numerical method like ADMM, which requires strict normalization~\cite{vial2022learning} or profiling~\cite{2ACE}.
The time is not negligible when the number of iterations is high.
Finally, $P(r,\theta)$ is not convex with regard to $S(x,y)$. 
Therefore, a general iterative algorithm can provide different solutions with different initial points and learning rates.
To acquire a good solution, multiple initial points and learning rates are often required.
Due to the large number of elements in $S(x,y)$, the number of initial points can grow too large to perform feasible computations.

\sssec{\name's Efficient Solver:}
We design a two-stepped phase retrieval approach to the problem. 
In the first step, we first treat the whole $\exp(-\bm{R}_r\bm{S})$, denoted by $\bm{C}$, to be the unknown, and solve $\bm{C}$ from $\bm{P}$ with constraints.
In the second step, we solve $\bm{S}$ from $\bm{C}$. 

\sssec{[Step 1] Solve $\bm{C}$ from $\bm{P}$: } Solving $\bm{C}$ is not intuitive due to the shape of $\bm{R}_x$, which is $N_rN_\theta\times N_{vr}N_\theta$, where $N_r$ is the number of rays and $N_{vr}$ is the number of virtual rays. 
The problem can be formulated as follows.
\begin{align}
    \begin{split}
        \min_{\bm{C}} \quad & \lVert \bm{P}-\bm{R}_x\bm{C}\rVert_2^2, \\
        \mathrm{s.t.} \quad & 0\leq\lvert\bm{C}\rvert\leq1, \text{~~~} \angle\bm{C}\leq 0.\\
    \end{split}
\end{align}

Solving $\bm{C}$ from $\bm{P}$ is a magnitude and phase constrained complex quadratic programming problem, which is non-convex in the complex domain.
Due to the definition, $\bm{R}_x$ is a rank-deficient matrix, as $N_{vr}\geq N_r$ to make sure that no two rays are identical.
Although the problem to solve $\bm{C}$ is condition-insufficient, i.e., there can be multiple solutions for $\bm{C}$, we note that the structure for $\bm{C}$ is special. 
As $\bm{C}$ represents the signal after attenuation of each sub-ray, the real part of the middle sub-rays will be generally smaller than the side sub-rays, since the object is usually placed in the middle. 
The imaginary part of the middle sub-rays will also be larger than the side due to the same reason.
To leverage the property and find a good solution for $\bm{C}$, we invert the problem into a new problem: finding a good approximation to the pseudo-inverse of $\bm{R}_x$, denoted by $\bm{R}^{-1}_x$, such that $\bm{R}^{-1}_x\bm{P}$ gives a good estimation of $\bm{C}$. 
To find a proper approximation of $\bm{R}^{-1}_x$, we first construct multiple synthetic values of $\bm{C}$ generated by synthetic values of $\bm{S}$. 
We then solve $\bm{R}^{-1}_x$ from the synthetic values of $\bm{C}$ and $\bm{P}$ by $\bm{R}^{-1}_x=\mathrm{pinv}(\bm{R}_x\bm{C})\bm{C}$, where $\mathrm{pinv}(\bm{R}_x\bm{C})$ is the Moore-Penrose pseudo-inverse of the synthetic $\bm{R}_x\bm{C}$.
Note that although $\bm{R}_x$ is a real matrix, $\bm{R}^{-1}_x$ is a complex matrix.
The value of $\bm{C}$ solved from $\bm{P}$ can then be inferred as follows: $\bm{C}=\bm{R}_x^{-1}\bm{P}$

\sssec{[Step 2] Solve $\bm{S}$ from $\bm{C}$}: After finding the the $\bm{C}$, the propagation constant map $\bm{S}$ can be solved through solving the following optimization problem.
\begin{align}\begin{split}
    \min_{\bm{S}} \quad & \left\lVert\bm{C}-\exp\left(-\bm{R}_r\bm{S}\right)\right\rVert_2^2+\mu\left\lVert\bm{S}\right\rVert_2^2,\\
    \mathrm{s.t.} \quad & \mathrm{Re}[\bm{S}]\geq 0, \text{~~~} \mathrm{Im}[\bm{S}]\geq 0.
    \end{split}
\end{align}

We can denote $\bm{C}$ as a combination of the amplitude and phase, which is $\bm{C}=\exp(-\bm{C}_r-j\bm{C}_\phi)$, where $\bm{C}_r\in\mathbb{R}^{N_{vr}N_\theta\times1}$ is the attenuation of $\bm{C}$, and $\bm{C}_\phi\in\mathbb{R}^{N_{vr}N_\theta\times1}$ is the phase of $\bm{C}$. The optimization can then be rewritten as follows.
\begin{align}\begin{split}
    \min_{\bm{S}} \quad & \left\lVert\exp(-\bm{C}_r-j\bm{C}_\phi)-\exp\left(-\bm{R}_r\bm{S}\right)\right\rVert_2^2+\mu\left\lVert\bm{S}\right\rVert_2^2,\\
    \mathrm{s.t.} \quad & \mathrm{Re}[\bm{S}]\geq 0,
    \text{~~~}  \mathrm{Im}[\bm{S}]\geq 0.
    \end{split}
    \label{eq:opt}
\end{align}

We next claim that solving the above optimization is equivalent to solving the real part and imaginary part of $\bm{S}$ independently, when the change inside $\bm{C}_\phi$ is not significant (less than $\pi$, so a flip does not happen), which holds in our case.

\sssec{\emph{Lemma 1: }} For $|\Delta\bm{C}_\phi| << \pi$, solving the optimization Eqn.~\ref{eq:opt} is equivalent to the following two LCQP problems with Tikhonov regularization: 
\begin{align}\begin{split}
    \min_{\bm{S}} \quad & \left\lVert\bm{C}_r-\bm{R}_r\bm{S}_r\right\rVert_2^2+\mu\left\lVert\bm{S}_r\right\rVert_2^2,\\
    \mathrm{s.t.} \quad & \bm{S}_r\geq 0,
    \end{split}\\
    \begin{split}
    \min_{\bm{S}} \quad & \left\lVert\bm{C}_\phi-\bm{R}_r\bm{S}_\phi\right\rVert_2^2+\mu\left\lVert\bm{S}_\phi\right\rVert_2^2,\\
    \mathrm{s.t.} \quad & \bm{S}_\phi\geq 0.
    \end{split}
\end{align}
Where $\bm{S}_r\in\mathbb{R}^{N_xN_y\times1}$ denotes the real part of $\bm{S}$, and $\bm{S}_\phi\in\mathbb{R}^{N_xN_y\times1}$ denotes the imaginary part of $\bm{S}$. \hfill $\square$

\noindent \textit{Proof Sketch: } The detailed proof is omitted here, but the basic process is to expand terms of the form $\lVert\bm{A}\rVert_2^2$ into $\bm{A}'\bm{A}$, where $\bm{A}'$ represents conjugate-transpose. 
The proof follows from setting the derivative of the Lagrangian dual to 0. \hfill $\square$

A solution for $\bm{S}$ can then be acquired by combining the real and imaginary part, i.e., $\bm{S}=\bm{S}_r+j\bm{S}_\phi$. 
Physically, $\bm{S}_r$ corresponds to the attenuation constant in the propagation constant, where $\bm{S}_\phi$ corresponds to the phase constant. 
The physical meaning also explains why solving $\bm{S}$ from $\bm{C}$ can be done independently with the real and imaginary parts, as the two constants are two orthogonal physical constants.

\begin{figure}
    \centering
    \subfigure[Real part w/o diffraction]{\includegraphics[width=0.22\linewidth]{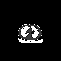}}
    \subfigure[Real part w/ diffraction]{\includegraphics[width=0.22\linewidth]{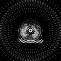}}
    \subfigure[Imag part w/o diffraction]{\includegraphics[width=0.22\linewidth]{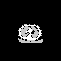}}
    \subfigure[Imag part w/ diffraction]{\includegraphics[width=0.22\linewidth]{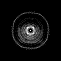}}\vspace{-10pt}
    \caption{Example of recovery result of a synthetic object. A brighter pixel corresponds to a larger value.}\vspace*{-10pt}
    \label{fig:diffraction_demo}
\end{figure}

\subsection{Recovery from Diffraction}
\label{subsec:unet}

To address diffraction, we take a data-driven approach rather than continuing the analytical modeling approach described above. We first discuss the motivation of this approach.

\sssec{Why learning rather than modeling?} The primary challenge is computational feasibility, i.e the diffraction model as formulated in Eq.~\eqref{eq:e_field_on_receiver_plane} and Eq.~\eqref{eq:power_diffraction} is non-convex with regard to the object. Hence, it is computationally intractable using traditional optimization methods.
Past systems that try to use optimization to recover from diffraction do not provide high enough resolution~\cite{mojabi2012microwave}.
Moreover, the shape and the size of the object are unknown, making it hard to directly predict and remove the diffraction pattern from the original measurement.
Although it is hard to solve diffraction from direct optimization, the diffraction pattern is a structured artifact.
For a propagation constant measurements, diffraction causes a (fictitious) hole in the middle of the object, and (fake) rings surrounding the object. 
Data-driven approaches such as  neural networks are proven to be able to remove structured artifacts in prior works~\cite{Prabhakara_2023}.
\name\ designs a U-Net to recover the propagation constant maps from diffraction.

\noindent\textbf{Understanding Diffraction artifacts: } We first analyze how diffraction causes errors in the synthetic aperture recovery process.
Figure~\ref{fig:diffraction_demo} shows a synthetic result whether we include a diffraction fraction with $\alpha=0.96$ in the received signal, as constructed by Eq.~\eqref{eq:power_combined}.
To understand the effect, imagine a round object being placed between the transmitter and the receiver.
If diffraction does not exist, a round shadow would be seen at the receiver plane based on the ray optics.
However, if we consider diffraction, we would not be seeing a full round shadow.
Instead, as can be imagined, a diffraction pattern forms at the receiver plane.
The diffraction pattern consists of a bright spot at the center (Arago spot/Poisson spot), as well as shadow patterns around the edge of the object (diffraction fringes).
As we only consider ray optics in synthetic aperture recovery, the solver would incorrectly process the Poisson spot as a hole in the middle of the object.
It would also erroneously consider the diffraction fringes as rings surrounding the object.
Together they constitute the artifact that we observe in Figure~\ref{fig:diffraction_demo}(b) and (d).


\noindent\textbf{Model selection: } We select U-Net architecture for our diffraction recovery task based on its proven effectiveness in wireless sensing applications, including signal segmentation~\cite{Contactless_seismocardiography}, denoising~\cite{Misaligned}, and super-resolution~\cite{Prabhakara_2023}. U-Net's encoder-decoder structure with skip connections is particularly well-suited for removing the structured diffraction artifacts in our propagation constant maps. The encoder path captures multi-scale features of diffraction patterns, while the decoder combined with skip connections preserves spatial information necessary for accurate reconstruction. Our implementation focuses on optimizing convolutional kernel sizes and network depth to capture the characteristic diffraction artifacts efficiently.


\begin{table}[t]
\centering
\caption{U-Net Layer-by-Layer Architecture}
\label{tab:unet_layers}
\begin{tabular}{lll}
\hline
\textbf{Block} & \textbf{Channels} & \textbf{Spatial Dim} \\
\hline
Encoder 1 & 2→16 & 64×64→32×32 \\
Encoder 2 & 16→32 & 32×32→16×16 \\
Encoder 3 & 32→64 & 16×16→8×8 \\
Encoder 4 & 64→128 & 8×8→4×4 \\
Encoder 5 & 128→256 & 4×4→2×2 \\
Encoder 6 & 256→512 & 2×2→1×1 \\
Bridge & 512→1024 & 1×1 \\
Decoder 6 & 1024+512→512 & 1×1→2×2 \\
Decoder 5 & 512+256→256 & 2×2→4×4 \\
Decoder 4 & 256+128→128 & 4×4→8×8 \\
Decoder 3 & 128+64→64 & 8×8→16×16 \\
Decoder 2 & 64+32→32 & 16×16→32×32 \\
Decoder 1 & 32+16→16 & 32×32→64×64 \\
Output & 16→2 & 64×64 \\
\hline
\end{tabular}
\vspace{-0.1in}
\end{table}

\noindent\textbf{Model design: } Our U-Net architecture consists of an encoder-decoder structure with skip connections specifically designed to handle diffraction artifacts in propagation constant maps. As shown in Table~\ref{tab:unet_layers}, we employ 6 encoder blocks that progressively reduce spatial dimensions while increasing feature channels from 2 to 512, followed by a bridge layer and 6 corresponding decoder blocks. Each encoder block contains two 3×3 convolutional layers with batch normalization and ReLU activation, followed by a 2×2 max pooling, while decoder blocks use transposed convolutions for upsampling. The deep 6-layer design reduces the spatial dimension to 1×1 at the bottleneck, which is critical for capturing global diffraction effects that manifest as structured artifacts across the entire image. The network processes 2-channel 64×64 images representing the real and imaginary parts of the propagation constant map, and produces output with identical dimensions after sigmoid activation.

\begin{figure}
    \centering
    \includegraphics[width=0.9\linewidth]{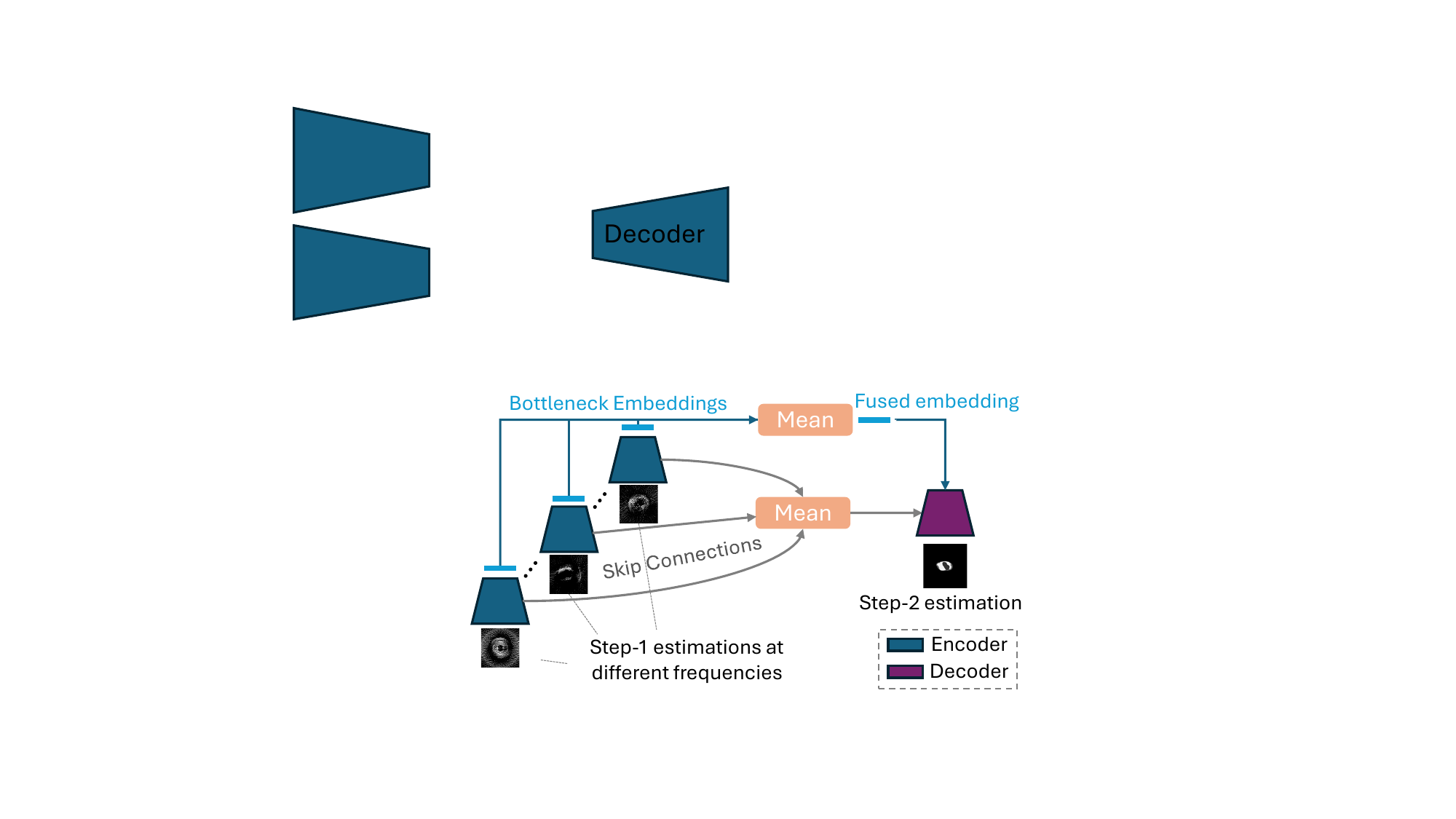}\vspace{-10pt}
    \caption{\name's diffraction removal neural network with U-Net architecture and multi-frequency fusion.}\vspace{-3pt}
    \label{fig:network}
\end{figure}

\noindent\textbf{Multi-frequency fusion: } With the design of the U-Net architecture which takes a single estimated image as input, we further seek to leverage the frequency diversity of the RF signals to reduce the diffraction artifacts. This design is based on a key observation --- even when the underlying target objects remain constant, the diffraction artifacts are different across frequencies. Specifically, we use the measurements from 11 frequencies spanning 5-6 GHz with 100 MHz spacing. As shown in Figure~\ref{fig:network}, the model processes the step-1 estimations from each frequency through separate encoder paths, then fuse the resulting embeddings by averaging at both the bottleneck and skip connections before feeding them into the decoder. This fusion strategy effectively suppresses frequency-dependent diffraction artifacts while preserving consistent structural information. The network learns to extract commonalities across frequency bands that represent the true object structure, while frequency-specific artifacts tend to average out. We note that the model weights in the encoder are shared across different paths and not frequency-dependent.  



\noindent\textbf{Data augmentation: } We employ two complementary augmentation strategies to enhance model robustness. First, we randomly sample 5 frequencies from the available 11 frequencies during each training epoch, forcing the network to generalize across different frequency combinations. Second, we augment both synthetic and real-world data through controlled rotation and displacement, where each augmentation applies offset values ($r_0$, $\theta_0$) to the raw $P(r, \theta)$ measurement matrix. Importantly, we perform optimization on each augmented input separately rather than augmenting the optimization outputs, as these approaches yield significantly different results. 

\begin{figure}
    \centering
    \includegraphics[width=0.95\linewidth]{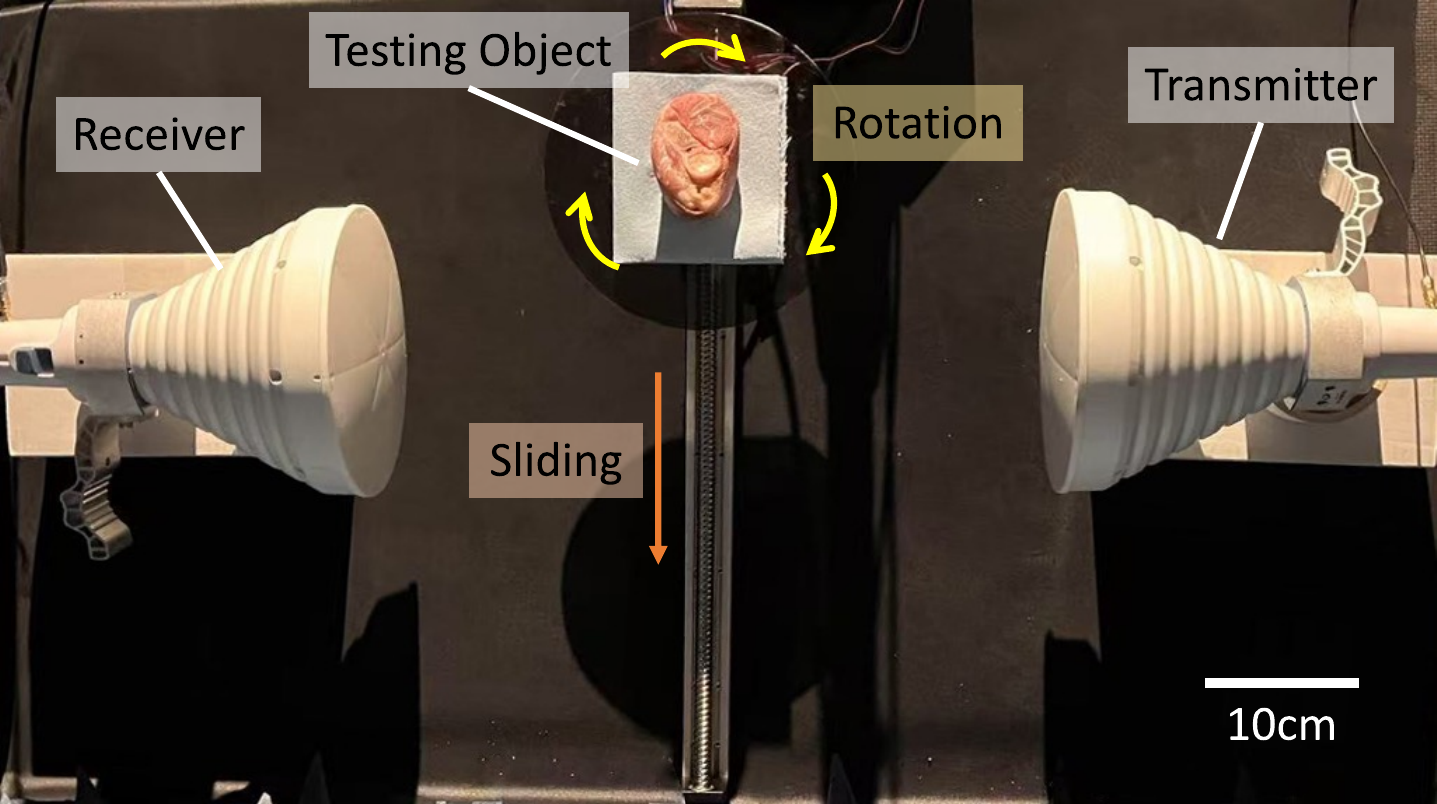}
    \caption{Experiment Platform: a pair of transceiver antennas, a sliding rail, and a rotation motor.}\vspace{-10pt}
    \label{fig:setup}
\end{figure}

\section{Implementation}\label{sec:implementation}

Figure~\ref{fig:setup} shows our hardware setup.
Data collection takes place inside an anechoic chamber to eliminate the effect of moving persons and objects.
We will discuss the effect of multipath in the result section.
We then discuss several factors that affect the imaging quality in the implementation.

\noindent\textbf{Frequency selection:} Choosing an appropriate operating frequency is relevant to the system performance. 
Lower frequency improves the wave's penetration, creating higher SNR, but increases antenna aperture and diffraction, decreasing resolution.
High frequency such as millimeter waves suffer from large attenuation through tissue (130~dB attenuation through a human body), as well as increased system cost for the need of high-speed electronics.
Our objective is to select the highest frequency in the ISM band that provides at most 10-30~dB attenuation across regular tissues.
The frequency is thus determined to be within the range of 5-6~GHz.
In our implementation, we take 11 uniform frequency points between 5-6~GHz to enrich the dataset diversity.

\noindent\textbf{Signal measurement device:} Signal measurement devices at the 5-6~GHz operating frequency span WiFi transceivers, software-defined radios (SDRs), vector network analyzers (VNAs), and oscilloscopes. 
We target a minimalist approach -- achieve a high imaging quality with a minimally-required commercial device.
With the proof-of-concept prototype, one can build more accurate imaging systems with higher-end wireless devices.
We use open-source LibreVNAs at a cost of several hundred dollars -- significantly less expensive than traditional VNAs. 
LibreVNA is a type of VNA that gives a high S21 measurement quality below 6~GHz.

\noindent\textbf{Rotation and displacement:} 
Antenna rotation/displacement is identical to object rotation/displacement.
We implement object rotation/displacement to reduce system bulkiness.
We use a FeeTech FT6335M step motor to control the rotation and a FUYU FSL40 linear rail slider to control the displacement. 
An acrylic plate of 1/4'' thickness and 15~cm diameter is used as the object holder.
The rotation step is 5 degrees for a total of 72 steps.
The sliding step is 2.5~mm for a total of 120 steps.
The data collection time is limited by the throughput of the VNA ($\sim$40 scans/second).
Therefore, the lower bound of the full data collection can be achieved within 216 seconds (3.5~minutes), and can be improved by switching to a VNA with a faster scan rate.

\noindent\textbf{Antennas:} Antennas affect the system performance. 
Ideally, we seek antennas with a high directivity to provide high SNR, and a small aperture to increase the performance of the synthetic aperture.
However, the Antenna Aperture Theorem states that the directivity is proportional to the aperture. 
We balance the directivity and aperture by choosing a directional antenna with an aperture smaller than the object size, the RFElements StarterHorn 30$^\circ$ USMA antenna with a 30$^\circ$ beam width with 18~dBi gain.


\noindent\textbf{Software implementation:} Our code will be open-sourced in publication.
Our grid space $S(x,y)$ is divided into a 61$\times$61-pixel grid constructing a 30~cm$\times$30~cm square area.
When evaluated on a desktop with Intel i9-12900KF, Nvidia RTX 3090Ti, and 64~GB memory, the optimization step takes 5.8~second for one image, and neural network inference takes less than 1~ms for one input.

\change{
\noindent\textbf{Dataset Preparation: }
We use 19 distinct samples to create the dataset for experimental evaluation. Among the 19 samples, there are 8 slices of pork forearms and 11 pork ribs, which contains different structures of bones and flesh.
We mark the ground truth of the samples by visually segmenting the slices of the samples into skin, muscle, and bones.
We then assign the real and imaginary parts of the RF propagation constant manually according to past measurements~\cite{gabriel1996compilation,gabriel1996dielectric1,gabriel1996dielectric2,gabriel1996dielectric3}.
The slices are of thickness ranging from 8~mm to 15~mm.
For slices that have minor variations in the front and back, we segment the cross-section of the front and back of the samples, and take the average as the ground truth.
We then create the ground truth of the same resolution (61$\times$61 pixels) as the output of \name\ for the convenience of training and validation.
}


\begin{figure*}
    \centering
    \includegraphics[width=0.9\linewidth]{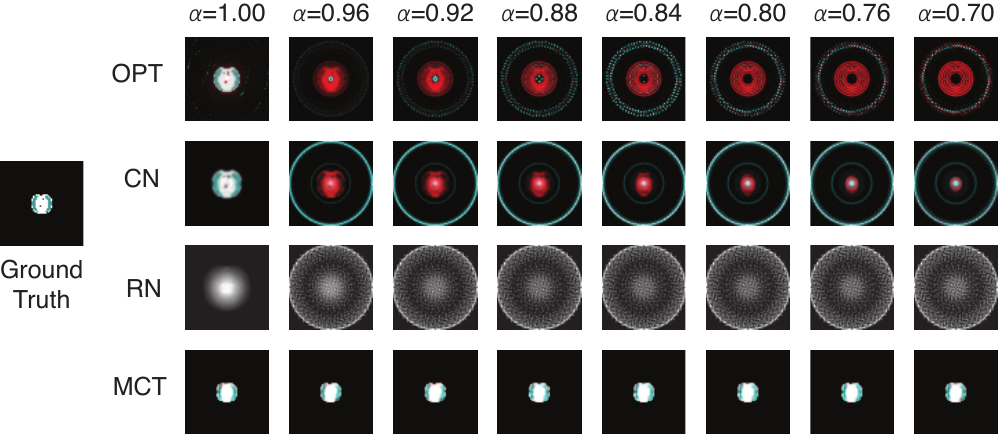}\vspace{-5pt}
    \caption{Impact of diffraction \change{through synthetic data}. OPT: Synthetic aperture optimization only. CN: Complex Newton's method. RN: X-ray Newton's method. MCT: Synthetic aperture optimization with neural network diffraction recovery.
    \change{Synthetic results show that RN only returns the shape and approximate size of the object even without diffraction, while other methods return the approximate shape and the inner structure of the object. OPT provides a higher resolution compared to CN. However, even with a small amount of diffraction, RN and CN provide a much worse recovery result, compared to OPT. OPT has resilience for large $\alpha$, but worsens when $\alpha$ decreases. MCT preserves high recovery quality with diffraction.}}\vspace{-5pt}
    \label{fig:diffraction-effect}
\end{figure*}


\section{Results}\label{sec:result}

We evaluate the performance of our system both through simulation and real-world measurements \change{(the majority of our results are real-world experimental -- from Sec. \ref{sec:expt} and beyond)}.

\noindent\textbf{Baseline:} Several algorithms are commonly adapted in microwave tomography~\cite{semenov2009microwave}. 
Newton's method~\cite{semenov2005microwave} is a well-evaluated method for in-tissue microwave imaging (noted as \textit{CN}).
It has a high resolution without diffraction, such as an experiment done in salt water~\cite{semenov2002three}, but has a near-wavelength resolution when diffraction exists~\cite{semenov2016electromagnetic}.
We replicate Newton's method by ray-optics modeling in the complex domain and adapt it as the baseline.
The real-valued ray-optical Newton's method (noted as \textit{RN}) is a derivation from X-ray CT method that has been adapted in microwave imaging as well~\cite{bolomey1990microwave}.
It is optimized for object detection as it amplifies the signal attenuation after passing the object, but is not sensitive to the inside of the object.

\noindent\textbf{Evaluation Metrics:} We aim to compare the structural similarity between the recovered image and the original segmentation. 
Therefore, we choose four evaluation metrics to reflect similarity: structural similarity index measure (SSIM, \textit{the larger the better}), peak signal-to-noise ratio (PSNR, \textit{the larger the better}), Mean Absolute Error (MAE, \textit{the smaller the better}), and Root Mean Squared Error (RMSE, \textit{the smaller the better}). 

\noindent\textbf{Visualization:} Since we are showing a complex image as result, we visualize the complex image by making the complex part being represented by difference colors in the RGB channel. 
We make the first (R) and second (G) channel to be the value of the real part of $S(x,y)$, normalized to 0-255, and the last (B) channel to be the value of the imaginary part, normalized to 0-255.

\change{
\noindent\textbf{Statistical Method:} To show the generality of \name, we use cross-validation during the evaluation. 
For cross-validation, we keep one dataset out as the test dataset, and the other datasets as the training and validation dataset (80\%-20\% training-validation ratio).
The keep-one-out testing is traversed among all datasets.
We show the mean value of the metrics in the bar plot, line plot, and CDF plot, and the error bars show the standard deviation of multiple trails among cross-validation.
}

\begin{figure}
    \centering
    \includegraphics[width=0.95\linewidth]{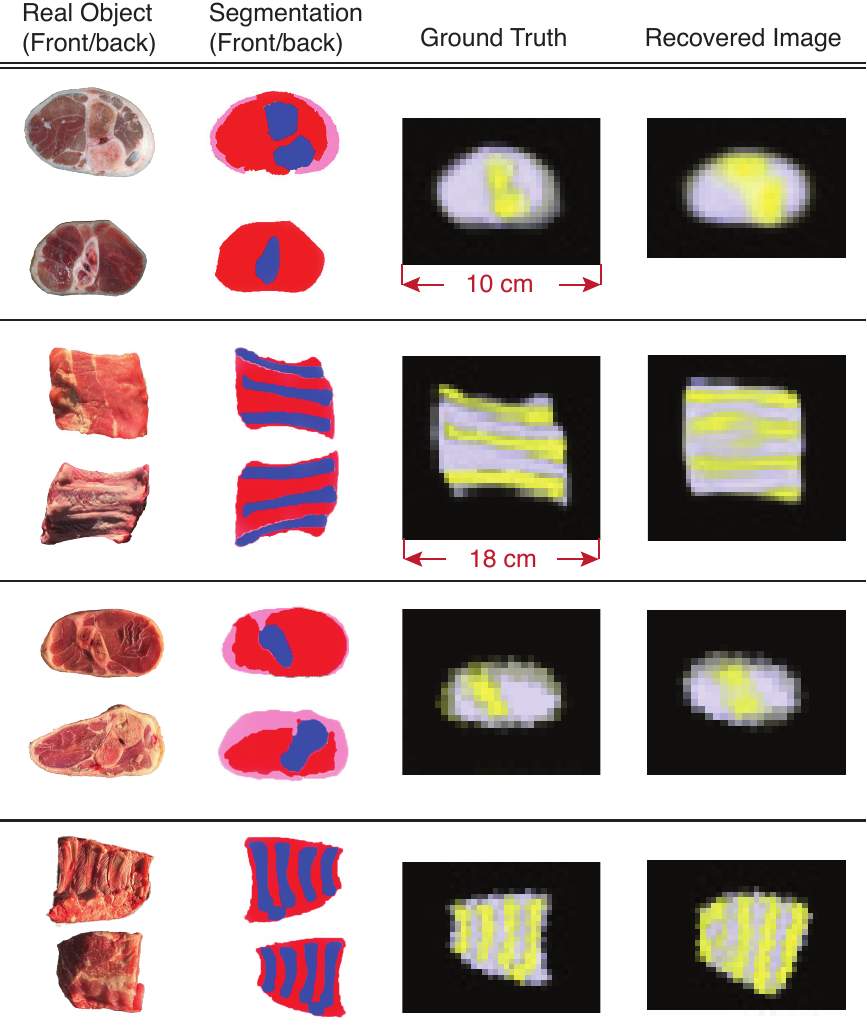}
    \caption{Examples of recovered images. In the segmentation: Blue corresponds to bone, Red corresponds to flesh. In the recovered image: Yellow corresponds to bone. White corresponds flesh. }\vspace{-5pt}
    \label{fig:image-example}
\end{figure}

\begin{figure*}
    \centering
    \includegraphics[width=0.9\linewidth]{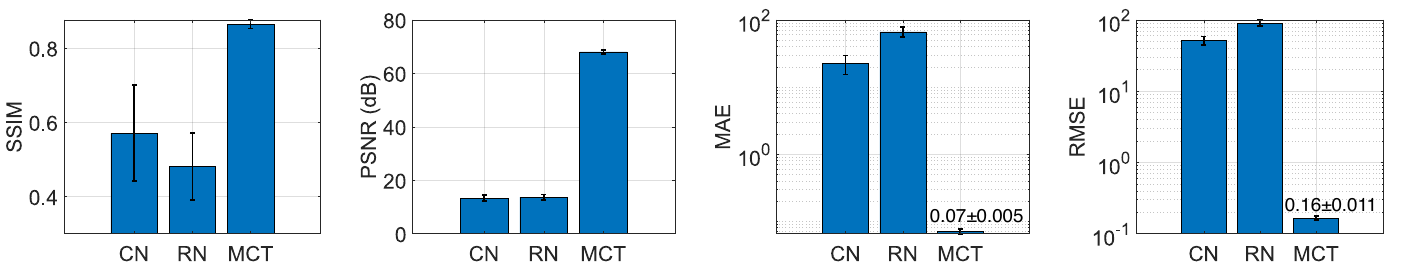}\vspace{-10pt}
    \caption{SSIM, PSNR, MAE, and RMSE values compared to baseline.}\vspace{-5pt}
    \label{fig:real-metrics}
\end{figure*}


    

\subsection{Impact of Diffraction}

As modeled in Eq.~\eqref{eq:power_combined}, we can control the strength of diffraction by controlling the parameter $\alpha$, where the higher the $\alpha$ is, the weaker the diffraction is. 
If $\alpha=1$, there is no diffraction.
It can be seen from Figure~\ref{fig:diffraction-effect} that the diffraction has a leading effect on the optimization-based recovery, for all algorithms.
It can be seen that without diffraction (ideal case), our optimization algorithm (OPT) provides the highest resolution of the object through the synthetic aperture method, while the complex Newton's method also provides some details, but the edge is more blurred.
In contrast, the real Newton's method only gives a hint of the object's existence without any details.
With small diffraction ($\alpha=0.96$), the edge and a rough inside of the object still can be inferred visually.
With stronger diffraction, the more part of the recovered image from optimization is taken by the diffraction artifact, becoming invisible.
However, the information about the object still exists and can be extracted through the neural network.
The last row of Figure~\ref{fig:diffraction-effect} shows the effect of the neural network over synthetic data.
It can be seen that the neural network is capable of recovering the shape of the original object without much loss of details.

\subsection{Real-World Experimental Data}\label{sec:expt}

\textbf{Visual result:}
We first show a visual recovery result in Figure~\ref{fig:image-example} of two forearm slices and two ribs to roughly demonstrate the capability of \name.
The forearm slices are roughly $5\times 8$~cm large, which is of a similar size to the wavelength, where the bones are much smaller than the wavelength.
It can be seen that \name\ is capable of recovering the correct size and shape of the object, while also predicting a correct position for the bones.
The ribs have different numbers of bones, and \name\ is capable of correctly counting them.
The result shows from an intuitive perspective of \name's sub-wavelength imaging capability inside the objects.

\noindent\textbf{Overall Performance: }
We show the performance of \name\ over CN and RN in Figure~\ref{fig:real-metrics}.
It can be seen that \name\ has a significantly better performance compared to CN and RN.
RN has an SSIM of 0.45, which can find the correct position of the object but gives no details of the object.
CN has an SSIM of nearly 0.6, which can find the correct position and give a rough estimate of the object's shape.
The PSNR of \name\ is higher as we can extract more details.
The recovery result of CN and RN follows the physical wavelength limit and results in other papers~\cite{9843919}.
A strong diffraction artifact existing inside the recovered image causes poor MAE and RMSE values.

\begin{figure}
    \centering
    \includegraphics[width=0.98\linewidth]{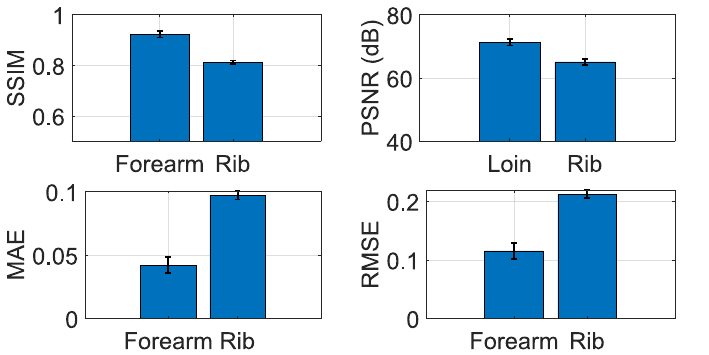}\vspace{-10pt}
    \caption{Different Object types affect imaging quality.}\vspace{-5pt}
    \label{fig:parts}
\end{figure}

\noindent\textbf{Object types: } 
We show the effect of different object types in Figure~\ref{fig:parts}.
It can be seen that the rib, where more complex inner structures exist, returns a worse value for all metrics. 
The SSIM sees a value difference of 0.1 and the PSNR sees a difference of 6~dB.
This is because the rib bones are much thinner than the forearm bones.
Therefore, while the forearm bones can always be imaged properly, the rib bones suffer from a bad resolution at the bone edges.
It can also be seen from Figure~\ref{fig:image-example} that the shape of the rib can not be recovered as well as the forearms, because the vertical structure of the rib is not homogeneous (the lower half is bone but the upper half is flesh).
Moreover, the edges of the rib bones are not as clear as the forearms.

\begin{figure}
    \centering
    \includegraphics[width=0.98\linewidth]{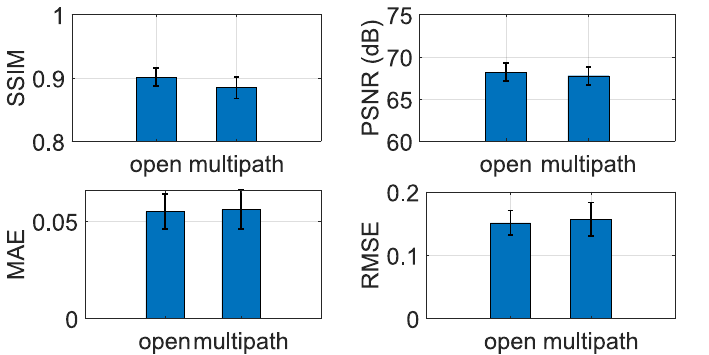}\vspace{-10pt}
    \caption{Multipath shows limited impact on image quality.}\vspace{-5pt}
    \label{fig:multipath}
\end{figure}

\noindent\textbf{Multipath environment: }
We assess the impact of multipath by deploying two reflectors 50~cm away from the data collection setup.
Figure~\ref{fig:multipath} shows the impact of multipath over the image quality metrics.
It can be seen that even with two strong static reflectors at a near corner of the setup, the multipath effect caused by such reflectors does not degrade much the image quality.
This is due to our selection of directional antennas and a signal pre-processing which cancels out the background.
The multipath result shows that \name\ can be deployed inside ordinary rooms.
However, we do acknowledge that randomly moving reflectors around the setup can cause more performance degradation due to the existence of the Doppler effect and randomly changing reflection signals.
But similar to CT scanners, we envision \name\ to be deployed in a relatively static environment.

\begin{figure}
    \centering
    \includegraphics[width=0.98\linewidth]{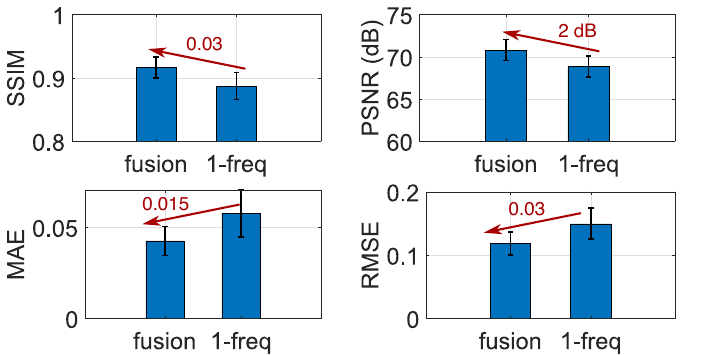}\vspace{-10pt}
    \caption{Frequency fusion shows improvement over single frequency.}\vspace{-5pt}
    \label{fig:freq_fusion}
\end{figure}

\noindent\textbf{Frequency fusion: } We show the improvement brought by frequency fusion in Figure~\ref{fig:freq_fusion}. 
With frequency fusion, more features of diffraction can be extracted, resulting in an improvement in the recovered image.
The SSIM sees a 0.03 improvement while the PSNR shows a 2~dB increase.
More significantly, the MAE and RMSE which correlates the image details, decrease by 0.010 (18\%) and 0.03 (20\%), respectively.
The result shows that frequency fusion can significantly improve the quality of the recovered image.

\begin{figure}
    \centering
    \vspace{-2pt}
    \subfigure[Changing number of ray displacements]
    {\includegraphics[width=0.95\linewidth]{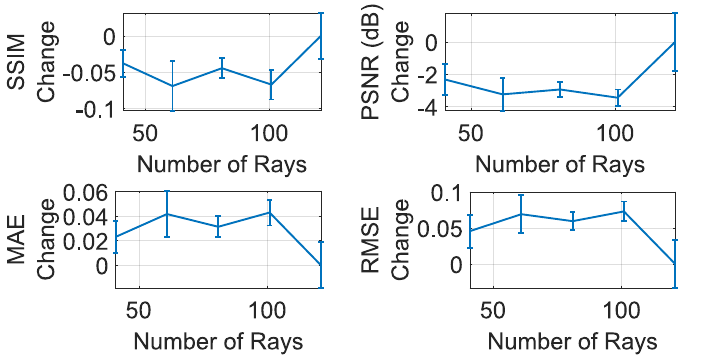}}\vspace{-10pt}
    \subfigure[Changing number of rotations]{\includegraphics[width=0.95\linewidth]{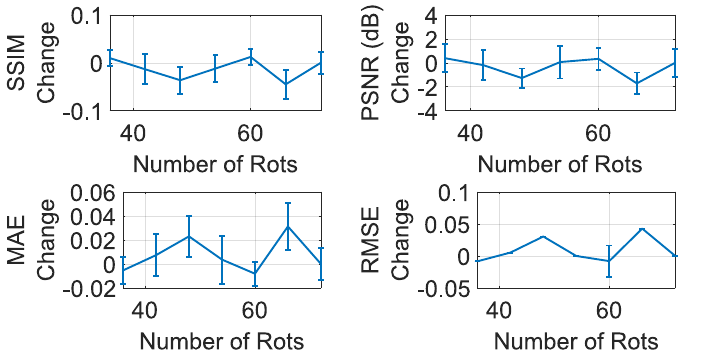}}\vspace{-10pt}
    \caption{Sensitivity to ray / rotation numbers (Rots) shows the impact of ray number to the system accuracy is larger than the impact of rotation numbers.}\vspace{-10pt}
    \label{fig:sensitivity}
\end{figure}

\noindent\textbf{Sensitivity over rotation and rays: } 
Since the recovered image has $N_xN_y$ pixels, theoretically we need the number of measurements $N_rN_\theta$ to be greater than the number of pixels.
However, as the recovered image is sparse, it is possible to find a recovery of the object image with fewer measurements.
To verify this, we decrease the number of measurements by subsampling the data.
First, we decrease the number of rays. 
As shown in Figure~\ref{fig:sensitivity}(a), it can be seen that the metrics worsen when the number of rays decreases.
This is because when the number of rays decreases, the intersection area between two rays increases.
As a result, the resolution for the synthetic aperture method decreases, causing detail in the recovered image to degrade.
We then decrease the number of rotations, where we can see a fluctuation in the metrics, showing that the system is not quite sensitive to the number of rotations, as shown in Figure~\ref{fig:sensitivity}(b).
This is because most objects are nearly central-symmetric, causing the rotational change to carry more repeated information about the object.

\begin{figure}
    \centering
    \subfigure[Segmentation error.]{\includegraphics[width=0.45\linewidth]{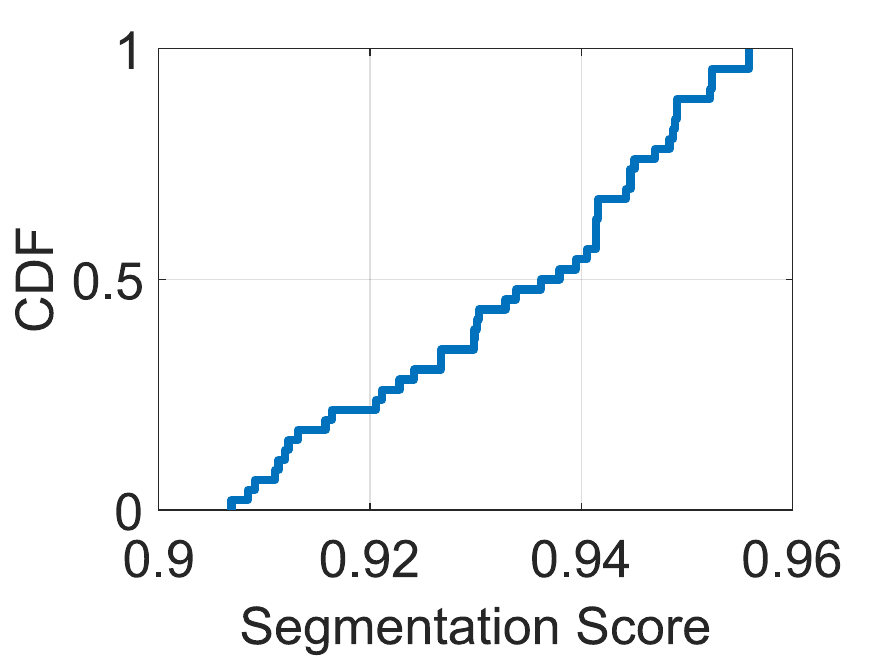}}\vspace{-5pt}
    \subfigure[Bone-muscle unit error.]{\includegraphics[width=0.45\linewidth]{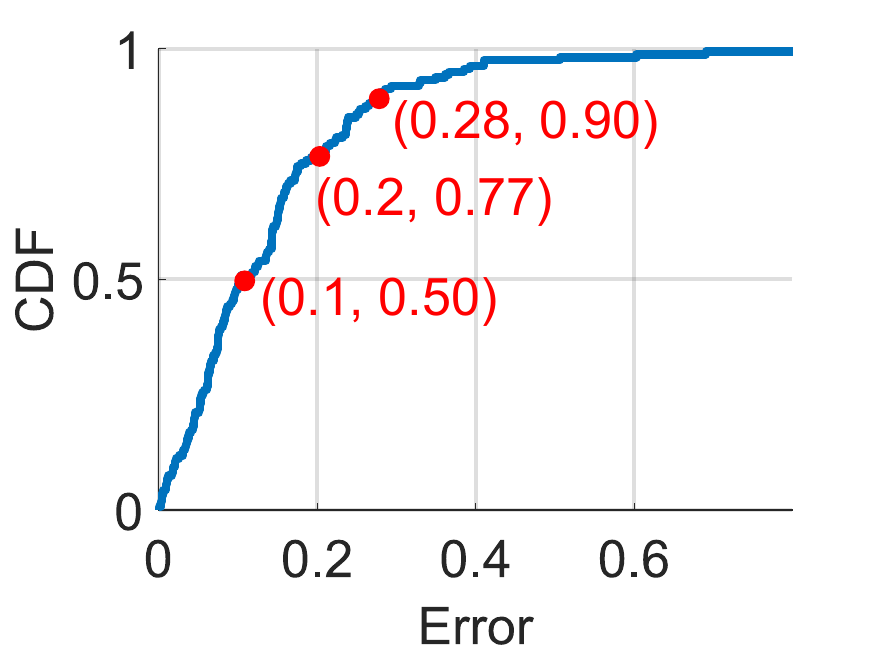}}\vspace{-5pt}
    \subfigure[Bone density.]{\includegraphics[width=0.45\linewidth]{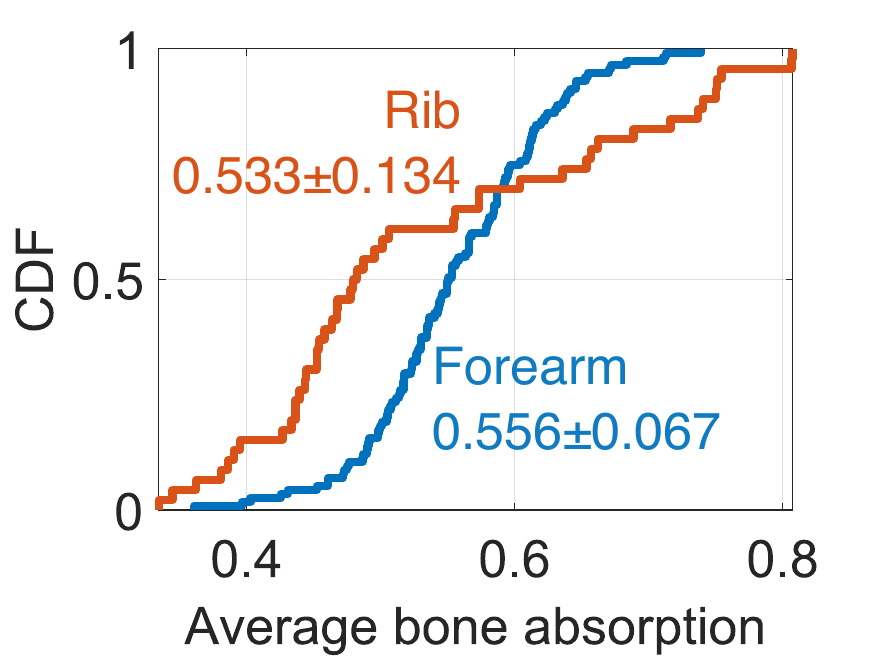}}\vspace{-5pt}
    \subfigure[Bone location.]{\includegraphics[width=0.45\linewidth]{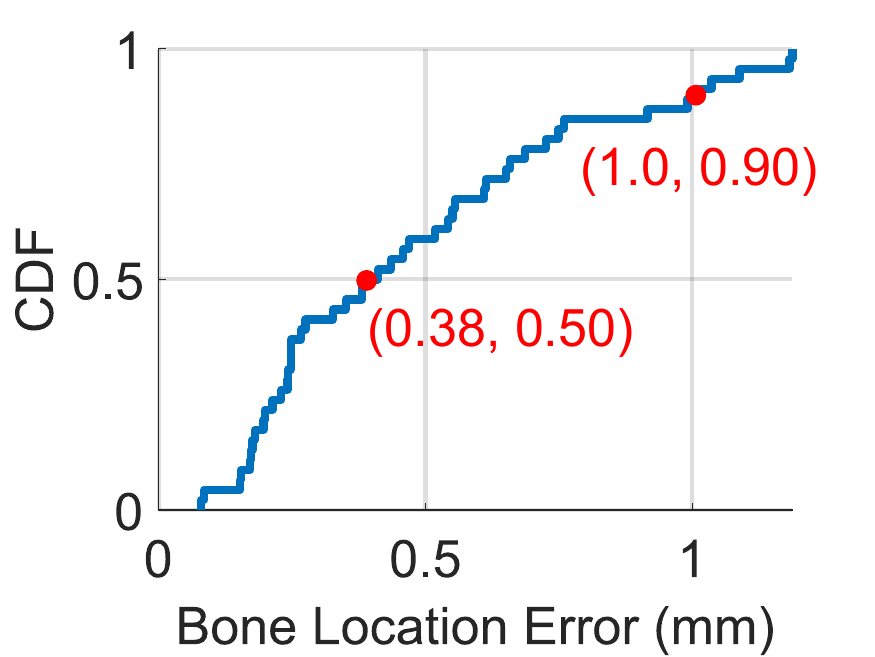}}\vspace{-5pt}
    \caption{We show the performance of \name\ over the four use cases. The results are computed by using the recovered image from \name, and we show the error CDF of the use cases.}\vspace{-10pt}
    \label{fig:use-cases}
\end{figure}

\subsection{Use cases}

We demonstrate four simple potential use cases for \name\ under the results we collected. 

\noindent\textbf{Bone segmentation:} The result from \name\ can be further processed by segmentation.
We use the general segment-anything~\cite{kirillov2023segment} toolbox and evaluate the segmentation accuracy by the first three optimal matching, representing the object, the bone, and the flesh.
The segmentation score is calculated by the overlapping ratio of the predictions over the ground truth.
As shown in Figure~\ref{fig:use-cases}(a), a 0.935 median segmentation score is achieved.

\noindent\textbf{Bone-muscle unit detection:} Bone-muscle unit refers to the fraction between bone and muscle, and is an important metric for accessing the bone health condition for cord health~\cite{cord_health}, strength~\cite{cord_health2}, and general health~\cite{TAGLIAFERRI201555}.
We calculate the fraction between bone and flesh by a simple thresholding method.
Figure~\ref{fig:use-cases}(b) shows the error for the bone-muscle unit, where a 26\% error can be achieved at 90\% cases.

\noindent\textbf{Correlation with bone density:} Bone density is an important measure for bone health detection. 
The bone density can be inferred by the average absorption rate within the bone area.
The bone mineral density of pig legs is greater than that of the pig ribs~\cite{bone_density}.
Figure~\ref{fig:use-cases}(c) shows a correlation, where the average value of the forearm is larger than the rib.

\noindent\textbf{Bone location detection:} \name\ has the potential for detecting a dislocation of bones by detecting the bone location.
As an example, we determine the bone location by calculating the center point (averaged x,y values) of the bones for the forearms.
The result in Figure~\ref{fig:use-cases}(d) shows that the median error is 0.38~mm and 90\% error is 1.0~mm.


\section{Discussion and Limitations}

\change{

\noindent\textbf{3D tomography: } This work focuses on 2D imaging of an object. 
The inner structure of a real-world 3-D object can change across different slices when taken along the vertical axis.
Past works have shown the possibility of migrating 2D imaging into 3D imaging by taking multiple 2D image slices at multiple altitudes and combining them to create a 3D image~\cite{kiemen2023tissue,mertzanidou20173d}.
Due to the unique propagation characteristic of RF, new algorithms can be created in future work to reconstruct a 3D model from 2D image slices.

\noindent\textbf{Resolution:} A typical early pathological change, such as early tumor or bone crack, is sub-cm level~\cite{peterson2005principles,chenevert1997monitoring}, which is the size of one pixel in the recovered image of \name.
Although the resolution of \name\ has been significantly improved compared to prior art, it still can be improved for better identification of changes and features on a small scale such as bone cracks.

\noindent\textbf{Combating Overfitting:} RF imaging through data-driven machine learning methods has been widely adapted and can suffer from overfitting~\cite{jiang2020towards, ji2023construct}. 
To combat overfitting, we have adapted several methods, including involving a more diverse dataset and data augmentation by rotation and shifting. 
We show our result in cross-dataset testing to prevent showing results from overfitting.
AI-driven imaging methods have also been validated and adapted in X-ray and MRI imaging~\cite{kambadakone2020artificial,wang2020deep,chen2022ai,lin2021artificial} to reduce the dosage, radiation exposure, and improve image quality.
As the dataset diversity of CT images is limited (e.g., different parts of the tissue), it is possible to create datasets that contain a large amount of microwave penetration profiles of different tissues to address overfitting.

\noindent\textbf{Power and Safety of Operation:} To improve the SNR, the operating power of our system is set at a relatively high level.
We use an emitting power of 1 Watt at 5-6 GHz, which is comparable to communication devices like WiFi~\cite{pozar2003regulations}.
Studies do not show direct evidence of harm to human of RF at such power levels~\cite{hardell2025critical,roosli2010systematic}.
According to experimental evidence, when using our current LibreVNA, the transmission power of 1 W reduces the impact of Additive White Gaussian noise (AWGN), while the colored noise in RF circuits dominates the error.

\noindent\textbf{Device:} Since the operation of \name\ requires the tissue and the environment to be static, we do not anticipate \name\ to be deployed on smart devices as in prior art on WiFi sensing. Instead, in its current form, \name\ is intended to serve as a bespoke medical imaging device (akin to traditional CT but modulo ionizing radiation).
\name\ requires a VNA to operate due to the stability and precision of VNA phase and attenuation data.
Compared to X-ray CT machines, MRI machines, and other microwave CT prototypes, \name\ has a much lower hardware cost ($\sim$\$2000 for VNA, antennas, and computer) and more lightweight. 
Therefore, we envision \name\ to have the potential of being more ubiquitous and accessible than X-ray CT and MRI, albeit imaging lower spatial resolution compared to these systems.

\noindent\textbf{Potential and Use Cases:} We propose \name\ as an early prototype and feasibility study to demonstrate using microwave to image inside the body. 
\name\ is of lower resolution compared to X-rays and MRIs, but avoids ionizing radiation and has a significantly lower operating cost.
For objects that have shown potential pathological changes under X-ray and MRI, a routine check happens every half a year, which may lead to significant progression of the disease.
In such cases, \name\ can be frequently applied to the specific region to monitor the progress of the change, and therefore, any treatment needed can be applied in a timely manner.
In addition, \name\ creates a new modality for sensing in addition to X-ray and MRI, which has the potential to distinguish pathological changes that are not readily visible under X-ray and MRI.
}

\section{Conclusion and Future Work}
This paper present \name, a microwave computed tomography (CT) system at radio frequency that can image 2D cross sections of bones at mm-resolution. \name\ achieves this through penetrative RF imaging at 5-6~GHz. \name's design contributes a novel penetration-based synthetic aperture algorithm that significantly increases imaging resolution. \name\ further mitigates the impact of diffraction through flesh using machine learning models. \name\ is extensively evaluated through meat models, demonstrating significant improvements in imaging resolution over prior art.

\vspace*{0.05in}\noindent\textbf{Future Work: } We leave addressing several important limitations of \name\ (as listed in Sec.~1) for future work: (1) We will explore ways to extend \name\ to 3D imaging; (2) We will explore solutions to optimize \name's computation to deliver faster imaging; (3) We will perform human trials of \name.


\newpage
\bibliographystyle{acm}
\bibliography{ref}

\end{document}